\documentclass[12pt]{article}

\usepackage{amsmath}
\usepackage{amssymb}
\usepackage{amsfonts}
\usepackage{amsthm}
\usepackage{graphicx,color}

\usepackage{latexsym}
\usepackage{color,soul} 
\usepackage{ulem}

\DeclareSymbolFont{matha}{OML}{txmi}{m}{it}
\DeclareMathSymbol{\varv}{\mathord}{matha}{118}

\usepackage{amsmath}
\usepackage{braket}
\usepackage{amssymb}
\usepackage{amsfonts}
\usepackage{latexsym}
\usepackage{graphicx}
\usepackage{latexsym}
\usepackage{xcolor}
\usepackage{epstopdf} 
\usepackage{indentfirst} 
\usepackage{cite}

\def\ln{\,\mbox{ln}\,}

\def\diag{\,\mbox{diag}\,}

\def\al{\alpha}
\def\be{\beta}
\def\ga{\gamma}
\def\Ga{\Gamma}
\def\de{\delta}

\def\th{\theta}

\def\ka{\kappa}

\def\La{\Lambda}

\def\ph{\varphi}

\def\om{\omega}
\def\Om{\Omega}
\def\na{\nabla}
\def\pa{\partial}

\renewcommand{\Re}{\,\mbox{Re}\,}
\renewcommand{\Im}{\,\mbox{Im}\,}

\def\beq{\begin{eqnarray}}
\def\eeq{\end{eqnarray}}
\newcommand{\eq}[1]{(\ref{#1})}
\newcommand{\n}[1]{\label{#1}}
\def\lap{\Delta}

\definecolor{Purple}{rgb}{0.4,0.3,0.}

\begin{document}

\begin{center}

{\Large
Weak-field limit and regular solutions in polynomial higher-derivative gravities
}
\vskip 8mm

{\large
Breno L. Giacchini$^{a}$
\ \ and \ \
Tib\'{e}rio de Paula Netto$^{b}$}

\end{center}
\vskip 1mm

\begin{center}
{\sl
(a) \ Centro Brasileiro de Pesquisas F\'{\i}sicas
\\
Rua Dr. Xavier Sigaud 150, Urca, 22290-180, Rio de Janeiro, RJ, Brazil
\vskip 3mm

(b) \ Departamento de F\'{\i}sica, \ ICE, \ Universidade Federal de Juiz de Fora, 
\\
36036-330 Juiz de Fora, \ MG, \ Brazil
\vskip 3mm
}
\vskip 2mm\vskip 2mm

{\sl E-mails:
\ \
breno@cbpf.br,
 \
tiberiop@fisica.ufjf.br}

\end{center}
\vskip 6mm

\begin{quotation}
\noindent
\textbf{Abstract.}
In the present work we show that, in the linear regime, gravity theories with more than four derivatives can have remarkable regularity properties if compared to their fourth-order counterpart. To this end, we derive the  expressions for the metric potentials associated to a pointlike mass in a general higher-order gravity model in the Newtonian limit. It is shown  that any polynomial model with at least six derivatives in both spin-$2$ and spin-$0$ sectors has regular curvature invariants. We also discuss the dynamical problem of the collapse of a small mass, considered as a spherical superposition of nonspinning gyratons. Similarly to the static case, for models with more than four derivatives the Kretschmann invariant is regular during the collapse of a thick null shell. We also verify the existence of the mass gap for the formation of mini black holes even if complex and/or degenerate poles are allowed, generalizing previous considerations on the subject and covering the case of Lee-Wick gravity. These interesting regularity properties of sixth- and higher-derivative models at the linear level reinforce the question of whether there can be nonsingular black holes in the full nonlinear model.
\vskip 3mm

\vskip 2mm

\vskip 2mm

\end{quotation}


\section{Introduction}
\label{Sec.1}

The problem of quantizing gravity is a long-standing one and many 
conceptually different approaches have been used to tackle it during the last almost 
five decades. One of the elements considered in the path towards quantum gravity are
the higher derivatives and the role they play in the
ultraviolet (UV) regime, where classical and quantum singularities show up.
Motivations for the introduction of curvature-squared terms in the action come already at 
semiclassical level, from the observation that the 
renormalization of quantum field theory on curved background requires 
 such higher-derivative terms~\cite{UtDW} (see also~\cite{book,ReviewSh} for a review).
Furthermore,
even though general relativity (GR) is not perturbatively renormalizable, its fourth-order counterpart
is~\cite{Stelle77}. Increasing the number of derivatives in the action can make the theory 
even more regular. For example, in the local theories with more than four derivatives it is possible to achieve
superrenormalizability~\cite{AsoreyLopezShapiro}. Indeed, the models with six derivatives have divergences only up to 3-loops, while in those with more than ten derivatives only 1-loop divergences remain. Moreover, in such models the $\beta$-functions are exact and gauge-independent.

The benefits 
that higher derivatives bring in what concerns renormalization,
however, come together with a serious
drawback regarding unitarity. Although it is possible to associate the
new degrees of freedom of the theory to positive-norm states in the Hilbert space,
some of them may carry negative energy~\cite{Stelle77,AsoreyLopezShapiro}. 
These so-called ghost states introduce instabilities
in the theory, with the possibility of a boundless vacuum decay via the emission of 
an arbitrary amount of energy in the form of gravitons. 
In such a
scenario it makes sense to study classical and quantum aspects of models which can offer insight on
how to deal with, \textit{e.g.}, the tension between renormalizability and unitarity,
or the most appropriate form of treating (or avoiding) ghosts and related instabilities%
\footnote{See, for example,~\cite{waveF1,waveF2,waveF3,waveF4,ABS-large} and references therein
for a discussion on some of the proposals and the difficulties they face.}.

In this regard, two models have been the subject of interesting investigations in  
recent years. The first one we mention is the Lee-Wick 
gravity~\cite{ModestoShapiro16,Modesto16}---see, \textit{e.g.},~\cite{ABS-large,Seesaw,Modesto-LWBH,Lens-LWBH,Modesto:2017hzl} for further developments and applications. 
This theory is defined by the Einstein-Hilbert action
enlarged by curvature-squared terms which contain polynomial functions of the d'Alembert operator,
such as $R_{\mu\nu} F_1(\Box)R^{\mu\nu}$ and $R F_2(\Box)R$. A general action of this type can be called polynomial higher-derivative gravity and was introduced in~\cite{AsoreyLopezShapiro};
the Lee-Wick gravity assumes, furthermore, that the polynomials $F_{i}$ are such that all
the massive poles of the propagator which correspond to ghost modes are complex. Hence, 
the physical spectrum of the 
theory contains the usual massless graviton and, possibly, a healthy massive scalar particle
(as the lightest scalar excitation is not a ghost~\cite{AsoreyLopezShapiro}).
The pairs of complex conjugate massive modes are understood as virtual ones only
and should decay to healthy particles. It was claimed that the presence of those complex poles 
do not violate the unitarity of the $S$-matrix if the Lee-Wick quantization prescription is 
used~\cite{ModestoShapiro16,Modesto16}.
Therefore, this could be a form of restoring the unitarity, weakening the tension between 
renormalizability and unitarity.

Another proposal for dealing with the problem of ghosts is to avoid them, at least at tree level,
by replacing the polynomials $F_i$ of the action by nonpolynomial functions of the 
d'Alembertian, which makes the theory nonlocal~\cite{Tseytlin95,Tomboulis,Modesto12,Maz12} (see also the earlier works~\cite{Krasnikov,Kuzmin}). It
is possible to choose these functions in such a manner that the theory propagator contains only the graviton pole%
\footnote{In the works~\cite{Tomboulis,Modesto12,Maz12,Krasnikov,Kuzmin} the nonlocality is introduced by the use of different types of functions, which may have particularities in what concerns the renomalizability properties of the model. 
For further considerations on quantum and formal classical aspects in nonlocal field theories see, \textit{e.g.}~\cite{ref0,ref1,ref2,ref3,ref4,ref5,ref6,ref7,ref8,ref9,ref10,ref11,ref12,ref13,ref14,ref15,ref16,ref17,ref18,ref19} and references therein.}, at $k^2=0$. 
Owed
to the absence of ghosts, this theory is sometimes called ghost-free gravity. As pointed out in~\cite{CountGhost}, however, quantum corrections may prompt the emergence of an infinite amount
of complex ghost poles. Therefore, the study of the Lee-Wick gravity may be useful also to the
better understanding and development of nonlocal UV extensions of GR.

The present work revisits two topics that have previously been investigated in the context of
local and nonlocal higher-derivative gravity models, namely, the nonrelativistic 
limit~\cite{ABS-large,Maz12,Newton-MNS,Newton-BLG,Quandt-Schmidt,Accioly-17,Head-On,EKM16} and
the collapse of small mass spherical shells~\cite{Frolov:Exp,Frolov:Poly}. 
Our focus is on general polynomial gravity, with a 
special attention given to the case of complex poles---Lee-Wick gravity---and 
also, for the sake of completeness, higher-order (degenerate) poles. In this sense, the results 
presented here both generalize and refine previous
considerations on the aforementioned topics, as we describe in what follows.

The presence of higher derivatives in the gravitational action tends to ameliorate both classical
and quantum divergences. The former can be viewed, \textit{e.g.}, on the Newtonian 
potential and on the effect of the gravitational collapse. 
The latter is related, as mentioned before, to the
(super)renormalizability of the theory. Since 1977 it is known that the fourth-derivative 
gravity is
renormalizable and has finite nonrelativistic potentials~\cite{Stelle77,Stelle78}. 
This relation was recently extended to superrenormalizable higher-order gravity theories with
real poles, which were shown to have a finite potential too~\cite{Newton-MNS}.
On the other hand, the introduction of higher derivatives only in the $R^2$-sector
of the theory results in a nonrenormalizable model
with a divergent (modified) Newtonian potential~\cite{Quandt-Schmidt}. 
These examples of simultaneous occurrence of classical and quantum singularities raised the 
question of whether there is a fundamental relation between 
them~\cite{CountGhost,Newton-MNS,Accioly13_riddle}.
The negative to this conjecture was given 
in~\cite{Newton-BLG}, where it was shown that the Newtonian singularity is canceled in all 
the polynomial gravity theories with at least one massive mode in each sector, which included 
Lee-Wick and also some nonrenormalizable models.

Nevertheless, the proof of the finiteness of the 
potential carried out in~\cite{Newton-BLG} was based on the calculation only of the terms which
give divergent contributions to the potential, and on the demonstration of an algebraic relation
between the poles of the propagator of the theory. In the Sec.~\ref{Sec.2} of the present work
we derive the expression for the weak-field metric potentials to all orders in $r$---including 
the case of degenerate poles---and obtain
an alternative verification of the cancellation of the Newtonian singularity. This simpler
demonstration is based on partial fraction decomposition and on the use of the heat kernel method for deriving gravitational potentials introduced in~\cite{Frolov:Poly}. 

Having the expression of the linearized metric for a pointlike source in a general local
higher-derivative gravity, it is possible to go beyond the analysis of the finiteness of the
potential and discuss the regularity of the curvature invariants. This is carried out also
in Sec.~\ref{Sec.2}, where we show that the metric is regular if and only if the model contains more than four 
derivatives in both the scalar and tensor sectors. This includes local superrenormalizable models and a wide class of Lee-Wick
gravities. Following the aforementioned parallel between quantum
and classical singularities~\cite{CountGhost,Newton-MNS,Newton-BLG,Accioly13_riddle}, one can say that GR is nonrenormalizable and has a divergent Newtonian
potential, fourth-order gravity is renormalizable and has a finite gravitational potential (but its curvature invariants diverge), and the higher-order gravities which are superrenormalizable have a complete regular nonrelativistic limit, {\it i.e.}, the metric potentials and the curvatures have no singularities.

In the Sec.~\ref{Sec.3}, the static solution found in the preceding section is used to 
obtain the metric associated to a nonspinning gyraton, which is an approximation to  an 
ultrarelativistic massive particle without angular momentum~\cite{Aich-Sexl,Frolov:2005in,Frolov:2005zq}. 
The procedure comprise applying a boost to the 
nonrelativistic metric and then taking the Penrose limit (see, {\it e.g.},~\cite{Frolov:book}). 
This turns out be an intermediate step to the analysis of the collapsing null shells.
The field generated by a nonspinning 
gyraton was derived in the context of the nonlocal ghost-free gravity 
in~\cite{Frolov:Exp}, and for the polynomial gravity with simple poles 
in~\cite{Frolov:Poly}. In the present work we show that this metric has the same 
small-distance behavior in all nontrivial polynomial gravity theories.
To conclude this section, some particular explicit examples are presented for 
the cases of complex and degenerate poles.

The collapse of small mass shells is analyzed in Secs. \ref{Sec.4} and~\ref{Sec.5} which,
respectively, discusses the case of an infinitesimally thin shell and of a shell with 
a finite thickness. 
By small mass we mean that we work only with linearized
equations for the gravitational field, in agreement to what was developed in the previous sections.
The interest in this scenario is the possibility of formation of mini black holes, 
\textit{e.g.}, owed to the collision of ultrarelativistic particles~\cite{Head-On}.
The formalism we follow was introduced in detail 
in~\cite{Frolov:Exp}, where it was applied to the ghost-free gravity.
It was later generalized in~\cite{Frolov:Poly}, where the case of polynomial models
with simple poles was considered. Our
extension of the latter work to general polynomial models verifies the conclusion
that there exists a mass gap
to the formation of mini black holes. 
The presence of a mass gap is typical in higher-derivative gravity models,
which is known since the 1980s~\cite{Frolov:Weyl}, and
 means that a black hole can only be formed if its mass is larger than a certain value. This is in contrast to
what happens in GR, where any mass can become a black hole, provided it is concentrated
in a sufficiently small region.

Also, in Secs.~\ref{Sec.4} and~\ref{Sec.5} we discuss the emergence of singularities during the
collapse of null shells within general polynomial gravities by analyzing the Kretschmann 
scalar $R_{\mu\nu\al\be}^2$. 
In particular, in Sec.~\ref{Sec.5} we show that the Kretschmann scalar for a collapsing thick null shell is regular for all models with more than four derivatives in the spin-2 sector. 
This completely characterizes the class of models for which $R_{\mu\nu\al\be}^2$ can have the logarithmic singularities found in~\cite{Frolov:Poly}.
Further discussion concerning similarities between local and nonlocal higher-derivative
gravity and extensions to the full nonlinear regime are carried out in Sec.~\ref{Sec.6}, where we also draw our conclusions.

Our sign conventions are $\,\eta_{\mu\nu} = \diag(-,+,+,+)\,$ for 
Minkowski spacetime metric and 
$\,{R^{\al}}_{\be\mu\nu} \,=\, \pa_\mu \Ga^{\al}_{\be\nu} - \dots\,$,
for the Riemann tensor. The Ricci tensor is defined by
$\,R_{\mu\nu} \,=\,{R^{\al}}_{\mu\al\nu}\,$.
Also, we use spatial distance and mass 
definitions such that $c\,=\,\hbar\,=\,1$.

\section{Newtonian limit}
\label{Sec.2}

In the static weak-field approximation we consider metric fluctuations around Minkowski spacetime
\beq
\n{mli}
g_{\mu\nu} \,=\, \eta_{\mu\nu} + h_{\mu\nu}
\,
\eeq
and work with the equations of motion at the linear level. The only relevant terms in the action which contribute for the linearized field equations are those of second order in the perturbation $h_{\mu\nu}$. 
Consequently, at the Newtonian limit
a general higher-derivative gravity model
can be reduced to the action
\beq
\n{act}
S_{grav} = \frac{1}{4 \ka} \int d^4 x \sqrt{-g} \,
\Big\{ 2  R 
+ R_{\mu\nu} \, F_1 (\Box)  \, R^{\mu\nu}
+\, R  F_2(\Box) \, R
\Big\}\,,
\eeq
where $\ka = 8 \pi G$ and $F_1$ and $F_2$ are functions of the d'Alembert operator. If $F_1$ and $F_2$ are nonzero polynomial functions, not necessarily of the same degree, we say it is a polynomial higher-derivative model\footnote{Note that the case of trivial polynomials,\textit{ i.e.}, $F_i=\text{const.}\neq 0$,  reduces to fourth-order theories; while the choice $F_1=F_2=0$ recovers GR.}. Otherwise, we say the theory is nonlocal.
Let us note that the term
$R_{\mu\nu\al\be} F_3 (\Box) R^{\mu\nu\al\be}$ is irrelevant for our
purposes since, by means of the Bianchi identities and integrations by parts,
one can prove that (see, \textit{e.g.},~\cite{AsoreyLopezShapiro})
\beq
\n{gb}
\int d^4 x \sqrt{-g} \,\Big\{
R_{\mu\nu\al\be}F_3 (\Box) R^{\mu\nu\al\be}
-4R_{\mu\nu}F_3 (\Box) R^{\mu\nu}
 + RF_3 (\Box) R\Big\} \,=\, O (R^3) \,=\,O(h^3)
\,.
\eeq
Hence, the effect of such Riemann-squared term can be reproduced, at the linear level, by a redefinition of the functions $F_1$ and $F_2$.

Performing the expansion \eq{mli}, the bilinear form of the action 
\eq{act} reads \cite{Maz12}
\beq
\n{bili} 
S^{(2)}_{grav} &=& \frac{1}{4\ka} \int d^4 x \,\bigg[
\frac{1}{2} h_{\mu\nu} \, a(\Box) \, \Box h^{\mu\nu} 
-\, \frac{1}{2} \, h \, c(\Box) \, \Box h + h \, c(\Box) \, \pa_\mu \pa_\nu h^{\mu\nu}
\nonumber
\\
&&
- \, h^\rho _\nu \, a(\Box) \, \pa_\rho \pa_\mu h^{\mu \nu}
 + 
\frac{1}{2} \, h^{\mu\nu} \, \left[ a(\Box) - c(\Box) \right]  \, \frac{1}{
\Box} \,
\pa_\mu \pa_\nu \pa_\rho \pa_\om h^{\rho \om}
\bigg]\,,
\eeq
where we introduced the condensate notations
\beq
\label{eq NN1}
a(\Box) &=& 1 + \frac12\, F_1(\Box) \, \Box \, , \\
\label{eq NN2}
c(\Box) &=&  1 - 2 F_2(\Box) \, \Box - \frac12\,F_1(\Box)\, \Box 
\,.
\eeq

The variational principle then yields the field equations,
\beq
\n{lieq}
& a (\Box) \, (\Box h_{\mu\nu} - \pa_\rho \pa_\mu h^\rho_\nu - \pa_\rho 
\pa_\nu h^\rho_\mu) 
 + \, c(\Box) \, (\eta_{\mu\nu} \pa_\rho \pa_\om h^{\rho\om} - \eta_{\mu\nu} 
\Box h
+ \pa_\mu \pa_\nu h) &
\nonumber
\\
& + \, \left[a(\Box) - c(\Box) \right] \, \dfrac{1}{\Box} \, \pa_\mu \pa_\nu \pa_\rho \pa_\om 
h^{\rho\om}
= \, - 2 \ka  \, T_{\mu\nu},&
\eeq 
where $T_{\mu\nu}$ is the energy-momentum tensor sourcing the field. 
As far as we are interested in a pointlike source, 
we assume
\beq
T_{\mu\nu} \,=\, \rho\, \de_\mu^0 \, \de_\nu^0
\,,
\label{Tmn-punti}
\eeq
where $\rho \,=\, m \, \de^3 ({\bf r})\,$ is the mass density.
In this case the metric can be written in the isotropic form
\beq
\label{m-New}
ds^2 \,=\, - (1+ 2 \ph) dt^2 + (1 - 2 \psi) (dx^2+dy^2+dz^2)
\,.
\eeq
Here $\ph = \ph(r)\,$ and $\psi = \psi(r)\,$ are the Newtonian
potentials and $r \,=\, \sqrt{x^2+y^2+z^2}\,$. The metric potentials
can be obtained by solving 
\beq
\label{met-2}
[a(\lap)-c(\lap)] \lap \ph + 2 c (\lap) \lap \psi &=& \ka \rho
\,,
\\
\label{eqmet-1}
[a (\lap)- 3 c(\lap)] [\lap \ph - 2 \lap \psi] &=& \ka  \rho
\, ,
\eeq
which are, respectively, the $00$-component and the trace of the
equations of motion~\eq{lieq}. Moreover, the substitution $\Box \mapsto \lap$
was implemented, as the metric is static.

Instead of solving the system above directly for $\ph$ and $\psi$, it is more
convenient to work with their linear combination in the form of
\beq
\label{Chi-Ome-Def}
\chi \equiv \ph + \psi \,  \qquad \text{and} \qquad \, \omega \equiv \ph - 2\psi \, .
\eeq
Once the equations are solved for $\chi$ and $\omega$ it is straightforward to obtain the original metric potentials via
\beq
\label{pot_ori}
\ph = \frac{1}{3} (2 \chi + \omega) \, , \qquad \psi = \frac{1}{3} (\chi - \omega) \, .
\eeq

The reason for working with $\chi$ and $\omega$ is threefold: first, the field equations for these new potentials
have a simple structure in terms of the functions $a$ and $c$. 
In fact, Eqs.~\eqref{met-2} 
and~\eqref{eqmet-1} are equivalent to
\beq
\label{eqx}
&&
a (\lap) \lap \chi \, = \, \ka  \rho \, , \\
&&
b(\lap) \lap \omega \, = \, - \ka \rho / 2 \, ,
\label{eqxB}
\eeq
where the function $b(z)$ is defined by%
\footnote{The multiplicative factor $1/2$ was introduced in order to have $b(0)=a(0)=1$. With this choice  the discussion carried out in the Appendix~A applies directly to both $a$ and $b$, simplifying the
considerations of Secs.~\ref{Sec.2.2} and~\ref{Sec.2.3}.}
\beq
\label{b-poly}
b(\lap) \equiv \frac{1}{2} \left[ 3 c(\lap)- a (\lap) \right] .
\eeq
Second, the functions $a$ and $b$ above correspond precisely to the terms which appear
in the propagator associated to the theory~\eqref{act}~\cite{VanNieuwenhuizen:1973fi},
\beq
G(k) \,=\, 
\frac{1}{k^2a(-k^2)} \, P^{(2)} - \frac{1}{2k^2b(-k^2)} \, P^{(0-s)}
\,,
\eeq
where $P^{(2)}$ and $P^{(0-s)}$ are, respectively, the spin-$2$ and spin-$0$ projection operators~(see, \textit{e.g.},~\cite{book}; tensorial indices and the terms which are gauge-dependent were omitted for simplicity).
Indeed, the roots of the equations $a(-k^2)=0$ and $b(-k^2)=0$
determine the massive poles of the propagator and, therefore, the (massive) spectrum of the model. In this spirit, Eq.~\eqref{Chi-Ome-Def} splits the metric potentials into the contributions owed to the spin-2 modes (through $\chi$) and to the scalar modes (via $\omega$). Based on this relation between the roots of the equations  $a(-k^2)=b(-k^2)=0$  and the poles of the propagator, throughout the present work we shall refer to these quantities as either ``roots'' or ``poles''.

The third motivation for working with the special combination in the form $\chi=\ph+\psi$ is that the potential $\chi$ turns out to be the relevant one for the collapse of the spherical null shell
(see discussion in~Sec.~\ref{Sec.3} and in~\cite{Frolov:Poly}). The situation resembles what occurs in the light bending~\cite{Bend-R2}. Qualitatively, this happens because in the ultrarelativistic limit the interaction between particles and the gravitational field is similar to that of photons.

\subsection{Heat kernel solution}
\label{Sec.2.1}

Equations~\eqref{eqx} and~\eqref{eqxB} have the very same structure, the only difference
being the operator function. From now on we assume that $F_1$ and $F_2$ are polynomial functions, as our interest in this work is on higher-derivative polynomial gravity. Then, $a$, $b$ and $c$ are also polynomials, but with different coefficients, and the equations for $\chi$ and $\om$ are essentially the same.
Therefore, we explicitly work out the solution for~\eqref{eqx}
and, \textit{mutatis mutandis}, write down the solution for~\eqref{eqxB}.
The solution for $\chi$ can be easily found by means of 
the heat kernel approach, based on the Laplace transformation, as
carried out in~\cite{Frolov:Poly}.

Indeed, introducing the Green's function for~\eq{eqx} via
\beq
\hat{H} \cdot \hat{G} \,=\, \hat{1}
\,,
\eeq
where 
\beq
\label{H}
\hat{H} \,=\, a(\lap) \lap
\,,
\eeq
we have the integral solution
\beq
\label{cGF}
\chi (x) \,=\, 8 \pi G 
\int d^3 x' G(x,x') \, \rho (x')
\,.
\eeq
Let us now
assume that the inverse $\hat{H}^{-1}(\lap)$ of the operator~\eq{H} can be 
written as the Laplace transform of some function 
$f(s)$, that is,
\beq 
\label{H-1}
H^{-1}(-\xi) \,=\, \int_0^\infty ds \, f(s)\, e^{-s \xi} 
\,.
\eeq
Then, the $x$-representation of the Green's function
$\hat{G}$ reduces to
\beq 
\n{Gxx}
G(x,x') \,=\, \int_0^\infty ds f(s) \, 
\langle x \,| \, e^{s \lap} \,|\,x' \rangle
\, ,
\eeq
where 
\beq
\langle x \,| \, e^{s \lap} \,|\,x' \rangle
\,=\, K(|x-x'|;s) 
\,=\, \frac{e^{-|x-x'|^2/4s}}{(4\pi s)^{3/2}}
\, 
\eeq
is the heat kernel of the Laplacian.
By choosing $x=r$ \ and \ $x'=0\,$, formula \eq{cGF}
simplifies to
\beq
\label{cHK}
\chi (r) \,=\, 8 \pi G m 
\int_0^\infty ds\, f(s)\, K(r;s)
\,.
\eeq

Particularizing for the higher-derivative model \eq{act}, according to the fundamental theorem of algebra we can write the polynomial $a(-\xi)$ in the factored form\footnote{The factors $m_i^{-2}$ must be introduced because Eq.~\eqref{eq NN1} requires $a(0) = 1 $. Analogous factors must be introduced for the polynomial $b(-\xi)$, as $b(0)=1$ by definition.}
\beq \label{poly_a}
a(-\xi) = \prod_{i=1}^{N} \left(\frac{ m_i^2 + \xi }{m_i^2} \right)^{\alpha_i} \, ,
\eeq
where $\xi = - m_i^2$ (with $i\in \lbrace 1,2,...,N\rbrace$) is a root of the equation $\, a(-\xi) = 0\, $ and $\alpha_i$ is its multiplicity. 
Notice that if $\mathcal{N}$ is the degree of $a(\lap)$---\textit{i.e.}, if there are $2(\mathcal{N}+1)$ derivatives in the spin-2 sector---then $\sum_{i=1}^N \alpha_i = \mathcal{N}$.
With the focus on general polynomial models, we shall not make any initial restriction on the complex or real nature of the quantities $m_i^2$, nor on their multiplicity.

The function $f(s)$ for the general higher-derivative gravity can be promptly obtained by substituting~\eqref{poly_a} into~\eqref{H-1} and inverting the Laplace transform using expansion in partial fractions~\cite{Grad}. The result is
\beq
\label{f(s)}
f(s) = -1 + \sum_{i=1}^N \sum_{j=1}^{\alpha_i}  \,  A_{i,j} \, s^{j-1} \, e^{- s m_i^2} \, ,
\eeq
where the coefficients $A_{i,j}$ are obtained from the comparison with $H^{-1}(-\xi)$ in terms of its partial fraction decomposition, namely,
\beq
\label{Aij}
A_{i,j} = \frac{-1}{(\alpha_i - j)! (j - 1)!} \, \frac{d^{\alpha_i -j}}{d\xi^{\alpha_i-j}} \frac{\left( \xi + m_i^2\right)^{\alpha_i}}{\xi a(-\xi)}  \Bigg|_{\xi = - m_i^2} \, .
\eeq
Also, for compactness of notation, it is useful to define the symbol $A_{i,j}$ for $j > \alpha_i$ by setting $A_{i,j> \alpha_i}\equiv 0$.

The potential $\chi$ can thus be evaluated by substituting~\eqref{f(s)} into~\eqref{cHK}, which gives
\beq
\label{Pot.Interm}
\chi(r)  = -\frac{2Gm}{r} 
 +  \frac{G m}{\sqrt{\pi}} \sum_{i=1}^N \sum_{j=1}^{\alpha_i} \, A_{i,j} \int_0^\infty ds\, s^{j-\frac{5}{2}} \,e^{- (s m_i^2  + r^2 /4s)}  \, , 
\eeq
where we assume that $\Re m_i^2 > 0$ for the integrals to converge.
Under the change of variables $\, s m_i^2 \mapsto s \,$ each of the above integrals becomes
\beq
\label{Integral}
I_i = \int_0^\infty ds\, s^{j-\frac{5}{2}} \,e^{- (s m_i^2  + r^2 /4s)} 
= (m_i^2)^{\frac{3}{2}-j} \int_\Gamma ds\, s^{j-\frac{5}{2}} \,e^{- (s  + m_i^2 r^2 /4s)} \, ,
\eeq
with the last integral being carried out along the line $\Gamma=\lbrace w \in \mathbb{C}: w=m_i^2 t, \, t \in \mathbb{R}^+\rbrace$. In the case of a real root $m_i^2$ the integration remains along the positive real axis, while for complex roots the integration line undergoes a rotation in the complex plane, but its points still satisfy $\Re w > 0$. However, the integrand $h(s)$ on the \textit{r.h.s.} of~\eqref{Integral} is an analytical function  with only a removable singularity at the origin, and which vanishes for $\vert s \vert \rightarrow \infty$. Therefore, the integral of $h(s)$ along the oriented contour $\Gamma_\varrho = [0,\varrho] \cup C_\varrho \cup \lbrace w \in \mathbb{C}: w= m_i^2 (\varrho- t), \, t \in (0,\varrho] \rbrace$, where $C_\varrho$ is the circumference arc of radius $\varrho$ connecting the points $w_1=\varrho$ and $w_2=m_i^2 \varrho$, is null. Taking the limit $\varrho \rightarrow \infty$, it follows that $\int_0^{\infty} h(s) ds  = \int_{\Gamma} h(s) ds$. We conclude that even in presence of complex roots $m_i^2$ it is possible to perform the integration along the positive real axis. Then,
\beq
I_i = (m_i^2)^{\frac{3}{2}-j} \int_0^\infty ds\, s^{j-\frac{5}{2}} \,e^{- (s  + m_i^2 r^2 /4s)} 
= 2 \left( \frac{r}{2 m_i} \right)^{j-\frac{3}{2}} K_{\frac{3}{2}-j}(m_i r )  \, ,
\eeq
where we chose the square root of $m_i^2$ with positive real part and recognized in the integral a representation of the modified Bessel function of the second kind $K_\nu$~\cite{Grad}.
Hence, the potential $\chi$ is given by
\beq
\label{Pot.General}
\chi(r) = -\frac{2Gm}{r} 
 +  \frac{2G m}{\sqrt{\pi}} \sum_{i=1}^N \sum_{j=1}^{\alpha_i} \, A_{i,j}  \, \left( \frac{r}{2 m_i} \right)^{j-\frac{3}{2}}  \, K_{j-\frac{3}{2}}(m_i r)  \, .
\eeq

In deriving this result it was assumed that $\Re m_i^2 > 0$ and $\Re m_i > 0$. The last assumption is physically justified by the requirement that the potential decays to zero at large distances, as well as to avoid tachyons on the model. The former assumption, however, is related to the heat kernel method used to solve~\eqref{eqx} and the premise that the operator $\hat{H}^{-1}$ has the form of~\eqref{H-1}. Actually, the solution~\eqref{Pot.General} also holds for the cases in which the polynomial $a(-\xi)$ has roots with $\vert\Im m_i\vert > \Re m_i > 0$, as the Bessel functions provide the analytical continuation of each term in~\eqref{Pot.Interm} viewed as function of an arbitrary $m_i^2$ with $\Re m_i > 0$.
We point out that it is possible to obtain the potential~\eqref{Pot.General}, even though with a longer calculation, directly for the general case of $\vert\arg m_i\vert < \pi/2$ by means of the Fourier transform method and using Basset's representation of the modified Bessel functions~\cite{Wat}.

The case of GR ($a \equiv 1$) is a trivial example of the previous formulas, as $f_{\text{GR}}(s) = -1$ and $\chi_{\text{GR}}(r)=-2Gmr^{-1}$. Another direct example is if $\, a(-\xi)=0 \,$ has only nondegenerate (ND)  roots. Then $\alpha_i = 1$ for all $i$, and $f(s)$ boils down to~\cite{Frolov:Poly}
\beq
\label{f_ND}
f_\text{ND}(s) = -1 + \sum_{i=1}^{N} e^{- s m_i^2} \prod_{j \neq i} \frac{m_j^2}{m_j^2 - m_i^2}  \, ,
\eeq
while the potential is given by~\cite{Frolov:Poly}
\beq
\label{Pot.NG}
\chi_\text{ND}(r) & = & -\frac{2Gm}{r} \bigg[ 1 - \sum_{i=1}^N  \,  e^{-m_i r} \prod_{j \neq i} \frac{m_j^2}{m_j^2 - m_i^2} \bigg] \, .
\eeq

Since the only assumption in finding the solution for $\chi$ was that it satisfied Eq.~\eqref{eqx}, one can write down the solution for $\omega$ which satisfies~\eqref{eqxB}. Let $\mathcal{N}^\prime$ be the degree of the polynomial $b(-\xi)$ and let $-m_i^{\prime 2}$ (with $i \in \lbrace 1,2,...,N^\prime\rbrace$) be the roots of the equation $b(-\xi)=0$, each of them with multiplicity $\alpha^\prime_i$. Then, the formula for $\omega(r)$ can be obtained by simply making the substitution $(\chi,A_{i,j},a,m,N,m_i,\alpha_i) \mapsto (\omega,A_{i,j}^\prime,b,-\frac{m}{2},N^\prime,m_i^\prime,\alpha_i^\prime)$ in Eqs.~\eqref{Aij} and~\eqref{Pot.General}.

In view of~\eqref{pot_ori}, the modified Newtonian potential $\ph$ for a general higher-derivative polynomial gravity is given by
\beq
\label{Phi-Gen}
\ph(r) &=& -\frac{Gm}{r} 
+ \frac{4}{3} \frac{G m}{\sqrt{\pi}} \sum_{i=1}^N \sum_{j=1}^{\alpha_i} \, A_{i,j}  \, \left( \frac{r}{2 m_i} \right)^{j-\frac{3}{2}}  K_{j-\frac{3}{2}}(m_i r)  
\nonumber
\\
&&
- \, \frac{1}{3} \frac{G m}{\sqrt{\pi}} \sum_{i=1}^{N^\prime} \sum_{j=1}^{\alpha_i^\prime} \, A_{i,j}^\prime  \, \left( \frac{r}{2 m_i^\prime} \right)^{j-\frac{3}{2}}  K_{j-\frac{3}{2}}(m_i^\prime r)  \, ,
\eeq
while $\psi$ reads
\beq
\label{Psi-Gen}
\psi(r) &=& -\frac{Gm}{r} 
+ \frac{2}{3} \frac{G m}{\sqrt{\pi}} \sum_{i=1}^N \sum_{j=1}^{\alpha_i} \, A_{i,j}  \, \left( \frac{r}{2 m_i} \right)^{j-\frac{3}{2}}  K_{j-\frac{3}{2}}(m_i r)  
\nonumber
\\
&&
+ \, \frac{1}{3} \frac{G m}{\sqrt{\pi}} \sum_{i=1}^{N^\prime} \sum_{j=1}^{\alpha_i^\prime} \, A_{i,j}^\prime  \, \left( \frac{r}{2 m_i^\prime} \right)^{j-\frac{3}{2}}  K_{j-\frac{3}{2}}(m_i^\prime r)  \, .
\eeq
As noted before, the quantities $m_i$ are the masses of the extra degrees of freedom with spin-$2$, while $m_i^\prime$ are related to the scalar ones. Moreover, the potentials are real despite the possibility of complex poles in the propagator. The cancellation of the imaginary part takes place because $K_n(\bar{z}) = \overline{K_n(z)}$ for $n \in \mathbb{R}$, and $A_{\bar{i},j} = \overline{A_{i,j}}$, where the subscript index $\bar{i}$ refers to the complex pole conjugate to $m_i^2$.

The general potential~\eqref{Phi-Gen} generalizes previous considerations found in the literature which  took into account real massive poles only in the scalar sector~\cite{Quandt-Schmidt}, or simple real poles~\cite{Newton-MNS} and simple complex poles~\cite{Newton-BLG,Frolov:Poly} in  scalar and tensor sectors. As noticed in~\cite{ABS-large,Newton-BLG,Quandt-Schmidt}, it is possible to obtain the potential for the case of degenerate poles by considering limits of the potential with only simple poles. This procedure may be ambiguous, however, when applied to poles with $\alpha_i > 2$. The formula~\eqref{Phi-Gen} clarifies the situation, as it explicitly allows for arbitrary multiplicity.

\subsection{Finiteness of the metric potentials}
\label{Sec.2.2}

If both $\chi$ and $\omega$ are finite, so are the metric potentials $\ph$ and $\psi$. As noticed in~\cite{Frolov:Poly}, if the roots of $a(-\xi)=0$ are all simple, then $\chi$ is finite. In what follows we use the general formula~\eqref{Pot.General} to show that $\chi$ is finite for an arbitrary nontrivial polynomial $a$ of the form~\eqref{eq NN1}. Using the similarity between the solution for $\chi$ and $\omega$, it then follows that these properties are valid also for $\omega$ defined by a nontrivial $b$ given by~\eqref{b-poly}. As a conclusion, if $a$ and $b$ have degree of at least one, then the potentials $\ph$ and $\psi$ are finite at $r=0$. This can be viewed as an explicit verification of the result obtained in~\cite{Newton-BLG}, where only the terms of order $r^{-1}$ were evaluated and the presence of degenerate poles was dealt with by the procedure of taking limits.

To this end, let us rewrite~\eqref{Pot.General} separating the terms for which $j > 3/2$:
\beq
\label{Pot.Gen-alt}
\chi(r)= -\frac{2Gm}{r} 
 + \frac{2G m}{\sqrt{\pi}} \sum_{i=1}^N \Bigg[ A_{i,1} \sqrt{\frac{2 m_i}{r}}  K_{-\frac{1}{2}}(m_i r) 
 + \sum_{j=2}^{\alpha_i}  A_{i,j}   \left( \frac{r}{2 m_i} \right)^{j-\frac{3}{2}}   K_{j-\frac{3}{2}}(m_i r) \Bigg] ,
\eeq
where the summation over $j \geq 2$ is considered only if $\alpha_i > 1$.
For $j \geq 2$ and small $r$ the functions $K_{j-\frac{3}{2}}(m_i r)$ behave like $r^{-j+3/2}$. Hence, all the terms with $j \geq 2$ are finite at $r=0$. It remains to check if the terms with $j=1$ manage to cancel the Newtonian singularity. Since
\beq
K_{\pm \frac{1}{2}}(z)= \sqrt{\frac{\pi}{2z}}e^{-z} \, ,
\eeq
the terms with $j=1$ have the form
\beq
 \frac{2G m}{\sqrt{\pi}} \sum_{i=1}^N  A_{i,1} \sqrt{\frac{2 m_i}{r}}  K_{-\frac{1}{2}}(m_i r) = \frac{2G m}{r} \sum_{i=1}^N  A_{i,1}  e^{-m_i r}  .
 \nonumber
\eeq
Therefore, the potential~\eqref{Pot.General} can be written as
\beq
\label{Pot.r=0}
\chi(r) = \frac{2Gm}{r} \left[ -1 + \sum_{i=1}^N  A_{i,1} \right]  +  \chi_0   +  \chi_1 r   + \, O(r^2) \, ,
\eeq
where $\chi_0$ and $\chi_1$ are constants. Using the identity
\beq
\sum_i  A_{i,1} = 1
\eeq
(see Eq.~\eqref{SumAi1} of the Appendix~A), it follows that the Newtonian singularity at $r=0$ is canceled by the higher-derivative correction terms, even in presence of complex and/or degenerate poles.

The same reasoning holds for the potential $\omega$, and therefore the metric potentials $\ph$ and $\psi$ are finite, verifying the result of~\cite{Newton-BLG}. The condition for the cancellation of the singularity of the potential is the presence of at least one massive mode in the spin-$2$ and in the spin-$0$ sectors. For example, if $F_1=0$ but $F_2 \neq 0$ then $\omega$ is finite but $\ph$ and $\psi$ are not~\cite{Quandt-Schmidt}.

\subsection{Regularity of the curvature invariants}
\label{Sec.2.3}

As it is well known, the finiteness of the potential is not enough to guarantee the regularity of the solution, as the curvature can still be singular. For a general metric in the form~\eqref{m-New}\textcolor{blue}, {\it e.g.}, the Kretschmann invariant 
\beq
R_{\mu\nu\al\be}^2 \,=\,
4 (\ph ''^2
+2\psi ''^2 )
+\frac{16}{r} \, \psi ' \psi ''
+\frac{8}{r^2}(
\ph '^2
+3 \psi '^2
)
\, ,
\eeq
clearly diverges if $\ph^\prime(0)$ and $\psi^\prime(0)$ are not zero.

In order to find more rigorously the conditions for having regular curvature invariants, let us assume that both metric potentials are finite and write
\beq
\ph(r) &=& \ph_0 + \ph_1 r + \ph_2 r^2 + \ph_3 r^3 + O(r^4)
\,,
\\
\psi(r) &=& \psi_0 + \psi_1 r + \psi_2 r^2 + \psi_3 r^3 + O(r^4)
\,.
\eeq
In terms of $\ph_n$ and $\psi_n$ the Kretschmann scalar reads
\beq
R_{\mu\nu\al\be}^2 =
\frac{8 (\ph_1^2+3 \psi_1^2)}{r^2}
+\frac{32 (\ph_1 \ph_2 + 4 \psi_1 \psi_2)}{r}
+
 48 \left( \ph_2^2 
+4 \psi_2^2
+ \ph_1 \ph_3
+5 \psi_1 \psi_3 \right)
+O(r) .
\eeq
Therefore, the invariant $R_{\mu\nu\al\be}^2$ is regular if, and only if, $\ph_1 = \psi_1 = 0$~\cite{Frolov:Poly,Buoninfante:2018b}. Actually, this is the same condition for the regularity of the set of curvature invariants\footnote{Note, however, that the invariants $R$ and $C_{\mu\nu\al\be}^2$ can be regular independently of the others as they depend, respectively, only on the scalar and on the tensor sectors.}:
\beq
R_{\mu\nu}^2 &=&
\frac{2(3 \ph_1^2-6 \ph_1 \psi_1 +11 \psi_1^2)}{r^2}
+ \frac{32 (\ph_1 \ph_2 - \ph_2 \psi_1
-\ph_1 \psi_2+ 4 \psi_1 \psi_2)}{r}
\nonumber
\\
&&
+ \,
12 \big[4 \ph_2^2 
+16 \psi_2^2
-8 \ph_2 \psi_2
+5 \ph_1 (\ph_3-\psi_3)
+  \psi_1( 21 \psi_3
-5 \ph_3) \big]
+{ O}(r) \,,
\eeq
\beq
R &=&
-\frac{4 \om_1}{r}
-12 \om_2
+ { O}(r)
\,,
\\
C_{\mu\nu\al\be}^2 &=&
\frac{4 \chi_1^2}{3 r^2}
- 8 \chi_1 \chi_3
+{ O}(r) \,,
\label{Weyl}
\eeq
where $C_{\mu\nu\al\be}$ is the Weyl tensor and
\beq
\chi_n = \ph_n + \psi_n 
\,,
\quad 
\om_n = \ph_n -2 \psi_n \,, 
\quad 
n \in \mathbb{N} \,.
\eeq

In this spirit, one may be tempted to ask whether the condition $\ph_1 = \psi_1 = 0$ is recurrent in higher-derivative gravity models. 
For example, there is a large class of non-local gravities that satisfy this condition when coupled to a $\delta$-source~\cite{Head-On,Buoninfante:2018b,Buoninfante:2018a} (see also~\cite{BreTib2} for more general non-local theories). On what concerns local models, the ones with only
fourth derivatives do not satisfy this condition~\cite{Stelle78,Stelle15PRL,Stelle15PRD}; however, it holds for the sixth-order gravity with a pair of complex poles~\cite{Modesto-LWBH}, and in~\cite{Holdom} it was given  general considerations supporting the conjecture that for theories with more than four derivatives one has $\ph_1 = \psi_1 = 0$. 
We here address a more direct answer to this question by explicitly showing 
which polynomial gravity models fulfil the
the conditions for having a regular metric at the linear regime.

To this end, let us extend to order $r$ the calculations of Sec.~\ref{Sec.2.2}. Using the general expression for the potential~\eqref{Pot.Gen-alt} and the series expansion of the modified Bessel functions for $j \geq 2$~\cite{Grad},
\beq
K_{j-\frac{3}{2}}(m_i r) =  \sqrt{\pi} \, e^{-m_i r}\sum_{k=0}^{j-2} \frac{(j+k-2)!}{k!(j-k-2)!(2m_i r)^{k+\frac{1}{2}}} \, , 
\nonumber
\eeq
it is not difficult to verify that the terms which contribute to order $r$ yield
\beq
\chi_1 = 2 G m \sum_{i=1}^N \Bigg\lbrace   \frac{A_{i,1} m_i^2}{2} - \frac{A_{i,2}}{2} 
 + \sum_{j=3}^\mathcal{N} \frac{A_{i,j}}{(4m_i^2)^{j-2}} \left[ \frac{(2j-5)!}{(j-3)!} - \frac{(2j-4)!}{2(j-2)!} \right]  \Bigg\rbrace  \, .
\eeq
But the term inside the summation over $j\geq 3$ is
\beq
\frac{(2j-5)!\left[ 2(j-2) - (2j-4)\right] }{2(j-2)!} = 0 \, .
\eeq
Thus,
\beq
\chi_1 = Gm (S_1 - S_2) \, ,
\eeq
where we define
\beq
\label{S_def}
S_1 = \sum_{i=1}^N A_{i,1} m_i^2 \, , \qquad S_2 = \sum_{i=1}^N A_{i,2} \, .
\eeq

In the Appendix~A we show that if the polynomial $a(-\xi)$ is of degree $\mathcal{N} > 1$, then $S_1 = S_2$ (see Eq.~\eqref{AppS1S2})---recall that $2(\mathcal{N} + 1)$ is the number of derivatives in the spin-2 sector of the action. It follows that for theories of order higher than four, the non-relativistic potential $\chi$ is not only finite, but it is also regular, \textit{i.e.}, $\chi_1=0$. On the other hand, for the case of $\mathcal{N} = 1$ with a root at $\xi=-m_1^2$ one has the trivial result $A_{1} = 1$ and $S_2=0$, which gives $\chi_1 = Gmm_1^2$.
This reasoning can be immediately extended to the potential $\om$, for which $\om_1 = -\frac{1}{2} Gmm_1^{\prime 2}$ if the polynomial $b(-\xi)$ is of order $\mathcal{N}^\prime = 1$, otherwise $\om_1 = 0$.

We conclude that the condition for the regularity of the curvature invariants\footnote{The calculation of the curvatures in this section was carried out using the GRTensor program (for analogous expressions in other parametrizations see, \textit{e.g.},~\cite{Frolov:Poly,Buoninfante:2018b,Buoninfante:2018a}). 
It is also possible to verify that under these conditions all individual components of the curvature tensors remain finite~\cite{BreTib3}.} is the presence of at least two massive modes (or one degenerate pole) in each of the spin-2 and the spin-0 sectors---which is equivalent to having $a$ and $b$ of degree higher than one\footnote{We point out that the effect of the regularization of the curvature can be viewed also in the polynomial theories as a regularization of the source in the Poisson equation for the metric potentials~\cite{BreTib2}.}. In other words, all higher-derivative theories defined by nonconstant polynomials $F_2$ and $F_1\neq - 3 F_2$ are regular at the Newtonian limit. In particular, this holds for the superrenormalizable local higher-derivative gravity models, including Lee-Wick models.

In this context, the only possibilities for having a singular solution for a point source in the Newtonian limit is to have $F_1(\Box) = \text{\textit{const.}}$ or $F_1(\Box)=-3F_2(\Box)$.
In the first case the spin-2 sector contains the massless pole corresponding to the graviton and, possibly, one massive (ghost) particle. In terms of the definition of $\pi$-regularity~\cite{Frolov:Poly}, which means that $\pi_1 = 0$ for a metric potential $\pi(r)$, we can say such a solution is not $\chi$-regular, but it could be $\om$-regular provided that $F_2(\Box) \sim \Box^p$ with $p \geq 1$. 
For the second case, \textit{i.e.}, if $F_1(\Box)=-3F_2(\Box)$, the solution is not $\omega$-regular.
Of course, for the solution to be regular it must be both $\chi$- and $\om$-regular.
In particular, Stelle's fourth order gravity is not regular at the Newtonian limit when coupled to a $\delta$-source~\cite{Stelle78,Holdom,Stelle15PRL,Stelle15PRD}, even though it can be $\ph$-regular for particular choices of parameters, namely, if $m_1^\prime=2m_1$.

\subsubsection{Small-$r$ conformally flat solutions}
\label{2.3.1}

In view of the Eq.~\eqref{Weyl}, it follows that a $\chi$-regular solution yields $C_{\mu\nu\al\be}^2 = 0$ at $r=0$. The components of the Weyl tensor read
\beq
C_{trtr} &=& \frac{1}{3}
\left(\chi '' - \frac{\chi '}{r}  \right)
\,,
\\
C_{t\th t\th} &=&  C_{r\th r \th} 
\,= \frac{C_{t\phi t\phi}}{\sin^2 \th}
=\, \frac{C_{r\phi r \phi}}{\sin^2 \th}
\,= - \frac12 \, r^2 C_{trtr}
\,,
\\
C_{\th\phi\th\phi} & = & - r^4 \sin^2 \th \, C_{trtr}
\,.
\eeq
As one can see, most of them do not contain terms
with powers of $r^{-1}$, which implies that if the potentials
are finite in the origin, the same is true for the corresponding
 components. The exception is the only
independent component, $C_{trtr}$. In fact,
\beq
C_{trtr} = - \frac{\chi_1}{3r} 
+ { O} (r)
\,.
\eeq
Thus, we conclude that this component is finite in $r=0$ only
for $\chi$-regular theories. In such a case, $\chi_1 = 0$ and the
components of the Weyl tensor tend to zero as $r \rightarrow 0$, which means
that the metric is approximately conformally flat near the origin.
This situation also holds in non-local higher-derivative gravity~\cite{Buoninfante:2018b,Buoninfante:2018a}.

\section{Ultrarelativistic limit}
\label{Sec.3}

Up to this point we restricted considerations to the Newtonian limit. In the following sections the weak-field potential $\chi$ will be used to discuss the emergence of a singularity in the collapse of null shells. As a first step towards the gravitational field of a collapsing shell, we shall obtain the field associated to an ultrarelativistic point-particle, which may be done by
the following procedure. First, we perform a Lorentz transformation 
into Eq.~\eq{m-New}, which yields the metric of a moving object 
with velocity $\be$.
Thereafter, we take the limit $\, \be \to 1\,$ while keeping 
the relativistic mass of the object fixed (Penrose limit),~\textit{i.e.},
\beq
\label{Pen}
\lim_{\ga \to \infty} ( \ga m ) \,=\, M 
\, ,
\eeq
being $M$ the mass of the ultrarelativistic particle
and $\, \ga = (1-\be^2)^{-1/2} \,$ the Lorentz factor.
The resultant metric 
corresponds to a nonspinning gyraton~\cite{Frolov:book}.

In order to apply this scheme to the solution found in the previous section, 
let us rewrite the  metric~\eq{m-New}
in the form
\beq
\label{ds2}
ds^2 \,=\, ds_0^2 + dh^2
\,,
\eeq
where
\beq
ds_0^2 \,=\, - dt^2 + dx^2 + dy^2 + dz^2
\eeq
is the flat spacetime metric and
\beq
dh^2 \,=\, - 2 \,[\ph dt^2 
+ \psi (dx^2 + dy^2 + dz^2) ]
\eeq
is the perturbation.

Now, consider a boost in the $x$-direction,
\beq
\n{tbo}
t = \ga \left(t' - \be\, x' \right)
,
\qquad
x = \ga \left(x' - \be\, t' \right)
.
\eeq
Introducing the null coordinates
$v=t'+x'$ \ and \
$u=t'-x'$, Eqs.~\eq{tbo}
read
\beq
\n{tuv}
t &=& \frac{\ga}{2} 
\left[\left(1-\be\right)v 
+ \left(1+\be\right)u  \right]
,
\\
\n{xuv}
x &=& \frac{\ga}{2} \left[\left(1-\be\right)v  
- \left(1+\be\right)u \right]
.
\eeq
Therefore, after applying the boost to the metric~\eqref{ds2} one gets
\beq
\n{fuv}
ds_0^2 \,=\, - 2 du dv + dy^2 + dz^2
\eeq
and
\beq
d h^2 = - \frac{\ga^2\left(\ph+\psi\right)}{2}
\left[\left(1-\be\right)^2 dv^2  
+ \left(1+\be\right)^2 du^2 
\right]
-\left(\ph-\psi\right) \, du\, dv
-2 \psi \,(dy^2 + dz^2)
\,.
\eeq
In the limit $\be \to 1$ the form of flat metric~\eq{fuv} 
remains unchanged, while the perturbation goes to
\beq
\label{gyr}
d h^2 = \Phi \, du^2 
\,, \qquad  \mbox{where}  \qquad
\Phi = - 2\, \lim_{\ga \to \infty} ( \ga^2 \chi ) 
\,.
\eeq
This shows, as mentioned before, that the 
dominant contribution in the 
ultrarelativistic limit comes from the special 
combination $\chi = \varphi+\psi$ of the metric potentials.
Owed to this fact, in this section and in Secs.~\ref{Sec.4} and~\ref{Sec.5} we restrict considerations to the spin-2 sector of the theory. In this spirit, when we refer to, \textit{e.g.}, ``models with more than four derivatives'' is must be understood that these derivatives are on the spin-2 sector.

The function $\Phi$ can be evaluated through~\eqref{gyr} 
by combining Eqs.~\eqref{cHK} and~\eqref{Pen} and
recalling that
\beq
\lim_{\ga \to \infty} 
\frac{\ga e^{-\ga^2 u^2 /4s}}{\sqrt{4\pi s}}
\,=\, \de(u) \, .
\eeq
Indeed, taking into account that $r^2 = \ga^2 u^2 + y^2 + z^2$ after the boost, it follows
\beq 
\Phi =
-4 G \lim_{\ga \to \infty} ( \ga m )  
 \int_0^\infty \frac{ds}{s}\, f(s) \, e^{-(y^2+z^2) /4s} \lim_{\ga \to \infty} 
\frac{\ga e^{-\ga^2 u^2 /4s}}{\sqrt{4\pi s}}
\, ,
\eeq
which can be written as
\beq 
\Phi \,=\,
-4 G M \, F(y^2+z^2) \, \de(u)
\,,
\eeq
where we defined the function $F\colon\mathbb{R}\rightarrow\mathbb{R}$ via
\beq
\label{Fdiv}
F(z) \,=\, \int_0^\infty \frac{ds}{s}\, f(s) \,e^{-z /4s}
\,.
\eeq
The integral~\eq{Fdiv} typically has an infrared divergence,
owed to the massless nature of the graviton. To overcome 
this problem one can introduce an infrared cutoff $\Om$ 
for large $s$. 
Any change in the cutoff parameter can be absorbed into a
redefinition of the coordinates. In other words, this ambiguity 
just reflects the freedom in the gauge choice.
Quantities with classical physical meaning, such as the curvature tensors,
do not depend on $\Om$ (for a more detailed exposition see, \textit{e.g.}, \cite{Frolov:Exp}).

For example, 
 $f(s) = -1$ in the case of GR, so that
\beq
F_\Om^{GR}(z) \,=\, - \int_0^{\Om^2} \frac{ds}{s}\, e^{-z /4s} \, = - E_1\left( \frac{z}{4\Om^2} \right) \, .
\eeq
Here $E_1(z)$ is the exponential integral function. As $\Om$ is a huge arbitrary cutoff, we assume $z \ll \Om^2$ and write
\beq
\label{F_GR}
F_\Om^{GR}(z) \,\approx\, \gamma + \ln \left( \frac{z}{\Om^2}\right) \, ,
\eeq
where $\gamma$ is the Euler-Mascheroni constant, and terms of order $z/\Om^{2}$ and higher were discarded.

For the general higher-derivative model~\eqref{act} the function $f(s)$ is given by Eq.~\eqref{f(s)}, which yields
\beq
F_\Om(z) = - E_1\left( \frac{z}{4\Om^2} \right) 
+ \sum_{i=1}^N \sum_{j=1}^{\alpha_i} A_{i,j} \int_0^\infty ds\, s^{j-2} \,e^{- (s m_i^2  + z /4s)} \, .
\eeq
By applying the same arguments used in Sec.~\ref{Sec.2.1} it is possible to express the function $F_\Om$ in terms of modified Bessel functions of the second kind,
\beq
\label{F_geral}
F_\Om(z) = - E_1\left( \frac{z}{4\Om^2} \right) 
+ 2 \sum_{i=1}^N \sum_{j=1}^{\alpha_i} A_{i,j} \left( \frac{\sqrt{z}}{2 m_i} \right)^{j-1} K_{j-1}(m_i \sqrt{z} )  .
\eeq

Before we present some explicit calculations for the cases of complex and degenerate poles, let us show a general property of the function in Eq.~\eqref{F_geral}. On the one hand, Eq.~\eqref{F_GR} shows that in GR $F_\Om^{GR}(z) \sim \ln z$ diverges as $z \rightarrow 0$. On the other hand, in~\cite{Frolov:Poly} it was shown that this divergence do not occur in the case of polynomial gravity with simple poles, because the leading terms of $F_\Om(z)$ for small $z$ are linear in $z$ or of the type $z \ln z$.
Now we prove that this feature is present also in the general polynomial theory. Indeed, for small arguments the modified Bessel functions of the second kind $K_n(z)$ ($n \in \mathbb{N}$) can be expanded as
\beq
K_0(z) & = & - \ln z + \frac{1}{4} z^2 (1 - \gamma + \ln 2) - \frac{1}{4} z^2 \ln z  
 +  c_0 + O(z^{4})  \, ,
\\
K_1(z) & = & \frac{1}{z} + \frac{z}{2} \left( \ln z + \gamma - \frac{1}{2} - \ln 2 \right)  +  O(z^{3})  \, ,
\\
K_n(z) & = &  \frac{(n-1)!}{2} \left( \frac{2}{z} \right)^{n} - \frac{(n-2)!}{2} \left( \frac{2}{z} \right)^{n-2} 
+  c_n + O(z^{-n+4}) \, , \quad  \text{for} \quad n \geq 2 \, ,
\eeq
where $c_i$ are constants and $\gamma$ is the Euler-Mascheroni constant.
Substituting these expressions in~\eqref{F_geral} and using~\eqref{SumAi1} it follows ($c^\prime$ is a new constant)
\beq
\label{F_expand1}
F(z) = - \frac{z}{4}  \left[ ( \ln z + 2 \gamma - 2 \ln 2 - 1)(S_1 - S_2) - S_1 + S \right] 
 +  c^\prime + O(z^2) \, .
\eeq
The constants $S_n$ are defined just like in~\eqref{S_def}, while $S$ is given by
\beq
\label{S_def2}
S = S_1^\prime - S_2^\prime + P_3 \, , \qquad
 S_n^\prime = \sum_{i=1}^N A_{i,n} (m_i^2)^{2-n} \ln m_i^2 \, , 
\qquad P_3 = \sum_{i=1}^N  \sum_{j=3}^{\mathcal{N}} \frac{(j-3)!}{ (m_i^2)^{j-2}} A_{i,j} \,  .
\eeq

Note that in any higher-derivative gravity model the singular term $\ln z$ which stems in GR (see~\eqref{F_GR}) is canceled by a specific combination of the contribution owed to each massive mode through $K_0(m_i \sqrt{z})$. This is a direct consequence of the cancellation of the Newtonian singularity discussed in Sec.~\ref{Sec.2.2} and in Refs.~\cite{Newton-MNS,Newton-BLG}. Also, while the constant $S_1^\prime$ is nontrivial for all higher-derivative polynomial models, the quantities $S_2$ and $S_2^\prime$ only appear if there is at least one pole with multiplicity equal or larger than 2, and $P_3$ is relevant only for models with at least one pole for which $\alpha_i \geq 3$---this justifies our choice for the subscript labels.

\subsection{Particular cases and examples}

To close this section let us consider some examples of the diversity of scenarios which occur in higher-derivative gravity. In particular, we present explicit calculations for the sixth-order gravity, which is the simplest model which admits complex or degenerate real poles.  We shall return to these examples in the next section, when analyzing the gravitational field of collapsing null shells.

\subsubsection{4th-order gravity}
\label{Ex.4th}

There is only one possible scenario: the equation $a(-\xi)=0$ has one real simple root at $\xi = - m_1^2$. Therefore, $S_1 = m_1^2$ and $S= m_1^2 \ln m_1^2$, so that
\beq
\label{F_4th}
F(z) &=& c^\prime  - \frac{z}{4}  \left(  \ln z + 2 \gamma - 2 \ln 2 - 2 + \ln m_1^2 \right)  m_1^2 
 + O(z^2) \, .
\eeq
As the other examples show, and in consonance with the discussion in Sec.~\ref{Sec.2.3}, this is the only case in which the small-$z$ expansion of $F(z)$ contains the term $z \ln z$. 

\subsubsection{Models with more than four derivatives}
\label{Ex.More}

For any model of order higher than four there is the identity $S_1 = S_2$ (see Eq.~\eqref{AppS1S2} of the Appendix~A). Hence, Eq.~\eqref{F_expand1} can be cast in a very simple form:
\beq
\label{F_expand2}
F(z) = c^\prime  - \frac{z}{4}  \left( S - S_2  \right)  + O(z^2) \, .
\eeq
This result is both a generalization and a simplification of the analogous expression derived in~\cite{Frolov:Poly}, as it accounts for the possibility of degenerate poles and also rules out the terms of the type $z\ln z$. 

\subsubsection{Nondegenerate models}
\label{Ex.Non-Deg}

The case of nondegenerate roots was investigated in Ref.~\cite{Frolov:Poly}. Here we show that our general considerations correctly reproduce this particular case.
If all the roots of $\,a(-\xi)=0\,$ are simple, then $\,\alpha_i = 1 \, \forall \, i \,$ and the general expression~\eqref{F_geral} for $F(z)$ reduces to~\cite{Frolov:Poly}
\beq
F(z) = - E_1\left( \frac{z}{4\Om^2} \right) + 2 \sum_{i=1}^\mathcal{N} K_{0}(m_i \sqrt{z} )  \prod_{j \neq i} \frac{m_j^2}{m_j^2 - m_i^2}  .
\eeq

Now, let us assume that $\mathcal{N} > 1$ (the case of $\mathcal{N} = 1$ was discussed in the Example~\ref{Ex.4th}). Inasmuch as all the roots are nondegenerate, it follows that $S_2=S_2^\prime=P_3=0$, whence $S=S_1^\prime$. Therefore, for small $z$ the function $F(z)$ behaves like
\beq
F(z) =  c^\prime  -  \frac{z}{4} S_1^\prime  +  O(z^2) \, .
\eeq

\subsubsection{Maximally degenerate models}

We say the higher-derivative model of order $\mathcal{N} > 1$ is maximally degenerate if the equation $a(-\xi) = 0$ has only one root at $\xi=-m_1^2$, with multiplicity $\mathcal{N}$. In such a case, the following relations are valid:
\beq
S_1^\prime = S_2^\prime  \, , \qquad S_2 = m_1^2 \, , \qquad
S = P_3 = \left\{ 
\begin{array}{l l}
0 \, ,  &  \text{if $\mathcal{N} = 2$,}\\
m_1^2 \sum_{j=3}^{\mathcal{N}} [(j-1)(j-2)]^{-1} \, , &  \text{if $\mathcal{N} > 2$,}\\
\end{array} \right .
\eeq
Thus, for small $z$ the function $F(z)$ can be written as
\beq
F(z)  = c^\prime + \frac{z}{4}  (  m_1^2 - P_3 ) +  O(z^2) \, .
\eeq

\subsubsection{6th-order gravity with simple poles}

For a pair of simple poles $m_1^2$ and $m_2^2$, Eqs.~\eqref{S_def} and~\eqref{S_def2} yield $S_2 = 0$ and $S_1^\prime \neq 0$. 
If these poles are simple and real the function $f(s)$ is given by
\beq
f(s) = -
1 + \frac{m_2^2}{m_2^2 - m_1^2} \, e^{- m_1^2 s} +  \,
\frac{m_1^2}{m_1^2 - m_2^2} \, e^{- m_2^2 s} \,,
\eeq
which yields, for small $z$,
\beq
\label{F_SR}
F(z) =  c^\prime +  \, \frac{m_1^2 m_2^2 \ln\left( \frac{m_1}{m_2} \right)}{2(m_1^2 - m_2^2)}   z \,  + O(z^2,z^2 \ln z) .
\eeq

In the case of two conjugate complex roots with $m_1=\al + i \be$ and $m_2=\al - i \be$, it follows
\beq
f(s) &=& -1 + \Big[ \cos(2\al\be s) 
 +  \frac{\al^2 - \be^2}{2\al\be} \sin(2\al\be s)\Big] e^{-s(\al^2-\be^2)} 
\eeq
and
\beq
\label{F_SC}
F(z) = c^\prime  +   \frac{(\al^2+\be^2)^2 }{4\al\be}\arctan\left( \frac{\be}{\al} \right) z 
+  O(z^2,z^2\ln z) .
\eeq

\subsubsection{6th-order gravity with degenerate poles}
\label{Sec.6th-degen.}

For degenerate real poles $m_1^2=m_2^2$ we have
\beq
f(s) = -1 + e^{-m_1^2 s} \left( 1 + m_1^2 s \right) .
\eeq
As the particular case of the $\mathcal{N} = 2$ maximally degenerate model, it holds $S_1^\prime = S_2^\prime = m_1^2 \ln m_1^2$, which gives $S=0$ and
\beq
F(z) & = &  \ln \left(\frac{z}{\Om^2} \right)
+ 2 K_0 \left( m_1 \sqrt{z} \right) 
+ m_1 \sqrt{z} \, K_1 \left( m_1 \sqrt{z} \right) 
\nonumber
\\
& = & c^\prime \, + \, \frac{z}{4} \,m_1^2 \, + \, O(z^2,z^2\ln z) \, .
\label{F_DR}
\eeq

We note that Eq.~\eqref{F_DR} can be obtained from the analogous equations for simple poles by taking the limit $m_2 \rightarrow m_1$ in~\eqref{F_SR}---or the limit $\be \rightarrow 0$ in~\eqref{F_SC}. 
While this procedure of taking the limit is simple to carry out in the case of two roots (see, \textit{e.g.},~\cite{ABS-large} for more examples), the situation might be not so clear if one is to consider a higher-order root. In such a case it is preferred to work with the general formula~\eqref{F_geral}, or~\eqref{F_expand2}, as discussed in Sec.~\ref{Sec.2}.

\section{Thin null shell collapse}
\label{Sec.4}

In this section we analyze the collapse of a null shell and the formation of mini black holes. Following Refs.~\cite{Frolov:Exp,Frolov:Poly}, we first consider a shell with vanishing thickness. For this case the Kretschmann curvature invariant is still singular, but this singularity is consequence of the nonphysical approximation of a infinitesimally thin shell.

The field associated to a thin null shell (or $\delta$-shell) can be obtained, at the linearized level, by the superposition of an infinite amount of gyratons spherically distributed and which pass through one given point $O$~\cite{Frolov:Exp}, which we take as the origin of the coordinate system. This point is the vertex of the null cone  representing the shell, so that for $t<0$ the shell is collapsing towards the apex $O$ and for $t>0$ it proceeds its expansion after the collapse. It can be shown that, outside the shell, the averaged metric perturbation $\langle dh^2\rangle$ resulting from this distribution of nonspinning gyratons is given by (see~\cite{Frolov:Exp} for a detailed derivation of this result)
\beq
\label{dhr}
\langle dh^2\rangle  = \frac{-2GM F(r^2-t^2)}{ r} \bigg[ \bigg( dt-\frac{t}{ r}dr \bigg)^2 
+ \frac{r^2-t^2}{
2}d\Omega^2 \bigg] \,, \qquad r \geqslant |t|\,, 
\eeq
where we use spherical coordinates, so that $d\Omega^2 = d\th^2 + \sin \th^2 d\phi^2$ is the metric of the unit sphere and
\beq
\label{Met_Thin}
ds^2 = -dt^2 + dr^2 + r^2d\Om^2 + \langle dh^2\rangle
\eeq
is the complete metric. Here $F(z)$ is defined by~\eqref{Fdiv}, as given by the metric~\eqref{gyr} associated to a single gyraton.

\subsection{Apparent horizon}

The formation of black holes  is closely related to the invariant
\beq
g \equiv (\nabla \varrho)^2=\frac{1}{4 f} \, g^{\mu\nu} \, \na_{\mu}  f \, \na_\nu f
\,,
\eeq
where $f = \varrho^2 \equiv g_{\theta\theta}$. Indeed, the points for which $g = 0$ correspond to an apparent horizon~\cite{FrolovNovikov}. If it happens that $g(t,r)$ is strictly positive then the collapsing shell generates no apparent horizon.

For the general metric~\eqref{Met_Thin} the invariant $g$ is given by~\cite{Frolov:Poly}
\beq
g = 1 - \frac{2GM}{r} \, q(r^2 - t^2)
\,,
\eeq
where 
\beq
q(z) \equiv z\, \frac{dF}{dz}(z)\, .
\eeq
If there is a positive constant $C$ such that
\beq
\frac{\vert q(r^2 - t^2) \vert}{r} < C \, ,
\eeq
then $g$ is positive anywhere provided that $M < (2GC)^{-1}$. Therefore, in order to show the existence of a mass gap to the formation of mini black holes one should verify that the function $r^{-1}q(r^2 - t^2)$ is bounded. In~\cite{Frolov:Poly} it was shown that for nondegenerate models there is the mass gap. In what follows we extend this result to the general polynomial model.

For $F(z)$ given by Eq.~\eqref{F_geral} we have
\beq
q(z) &=& 1 -  \sqrt{z} \, \sum_{i=1}^N A_{i,1}\,  m_i \, K_1(m_i \sqrt{z}) 
\nonumber
\\
&&  +\, 2 \sum_{i=1}^N \sum_{j=2}^{\alpha_i} A_{i,j} \, \left( \frac{\sqrt{z}}{2 m_i} \right) ^{j-1} \Big[ (j-1) K_{j-1}(m_i \sqrt{z}) 
- \frac{m_i \sqrt{z}}{2} K_j(m_i \sqrt{z}) \Big].
\eeq
As a finite sum of continuous functions defined for all $z \in \mathbb{R}^+$, $q(z)$ is also continuous. Hence, if $q(z)$ has any singularity it can only take place for large or small $z$. The former divergence does not occur, because the functions $K_{j}(z)$ decay exponentially as $|z| \rightarrow \infty$, in such a way that $q(z) \rightarrow 1$ as $z \rightarrow \infty$. On the other hand, assuming $\mathcal{N}>1$,
 for small arguments one has
\beq
\label{q-small}
q(z) =  - \frac{z}{4} \, \left(  S - S_2  \right) \, + \, O(z^2) \, ,
\eeq
whence $q(z) \rightarrow 0$ as $z \rightarrow 0$. Being the asymptotic limits finite, it follows that $q(z)$ is bounded. Now let us analyze the function $r^{-1}q(r^2 - t^2)$. The function $r^{-1}$ is continuous, it vanishes for large $r$ and only diverges as $r \rightarrow 0$. In this regime, however, $q(r^2 - t^2)$ dominates over $r^{-1}$, since $|t| < r$ outside the shell implies in $r^2 - t^2 < r^2$. Thus,
\beq
\lim_{r\rightarrow 0} \frac{\vert q(r^2 - t^2) \vert}{r} = 0 \, .
\eeq
Similar analysis can be applied for the case $\mathcal{N} = 1$, with the same result~\cite{Frolov:Poly}. We conclude that  $r^{-1}q(r^2 - t^2)$ is bounded for general polynomial gravity models, which implies in the existence of the mass gap for the formation of mini black holes. The size of the gap depends on the scale $\lambda = \max_i \lbrace m_i^{-1} \rbrace$ defined by the massive excitations of the model; such scale could be affected by a gravitational seesaw-like mechanism as discussed in~\cite{Seesaw} (see also~\cite{ABS-large,MG14} for experimental bounds on $\lambda$).

\begin{figure} [h]
\centering
\includegraphics[scale=0.75]{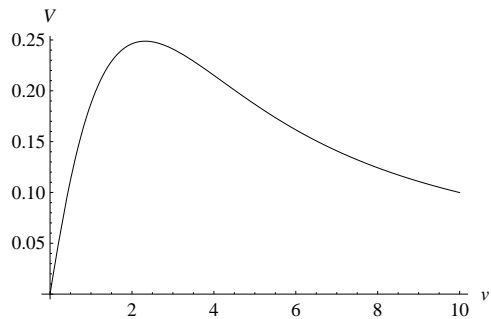}
\caption{Graph of $V(\varv)$ for the sixth-order gravity with degenerate roots.}
\label{Fig1}
\end{figure}

To give an example of an explicit calculation, consider the sixth-order gravity with degenerate poles
discussed in Sec.~\ref{Sec.6th-degen.}, for which
\beq
q(z) = 1 - \frac{m_1 \sqrt{z}}{2} K_1(m_1 \sqrt{z}) 
-  \frac{m_1^2 z}{4} [K_0(m_1 \sqrt{z}) + K_2(m_1 \sqrt{z})] \, .
\eeq
Following~\cite{Frolov:Poly} we put $\beta^2 \equiv 1 - t^2 r^{-2}$ and $\varv \equiv m_1 \beta r$, so that $r^{-1}q(r^2 - t^2) =   m_1 \beta V(\varv)$, with
\beq
V(\varv) =  \frac{1}{\varv} - \frac{1}{2} K_1(\varv) - \frac{\varv}{4} \left[K_0(\varv) + K_2(\varv)\right]  \, .
\eeq
The function $V(\varv)$ is positive and reaches its maximum of about 0.249 at $\varv \approx 2.324$ (see Fig.~\ref{Fig1}). Thus,
\beq
\frac{2GM}{r}  q(r^2 - t^2)  =  2GM m_1 \beta V(\varv) \lesssim   0.5 \,  GM m_1  ,
\eeq
as outside the shell the parameter $\beta$ ranges in the interval $(0,1)$. Therefore, if $M \lesssim 2 (G m_1)^{-1}$ the collapse does not result in a black hole.

\subsection{Kretschmann scalar}

Even though there is a mass gap for the mini black hole formation in the general higher-derivative gravity, the Kretschmann invariant is not regular at $r=0$. Indeed, for a metric in the form~\eqref{Met_Thin} it is given by~\cite{Frolov:Exp,Frolov:Poly}
\beq
R_{\mu\nu\al\be}^2 = \frac{48 G^2 M^2}{r^6} \, Q(r^2 - t^2) \, , \quad 
 \text{where} \quad Q(z) \equiv 2 z^2 q'^2 - 2 z q q' + q^2 \, 
\eeq
and primes denote differentiation with respect to the argument $z$. For $\mathcal{N} > 1$ and small arguments, $q(z)$ is given by~\eqref{q-small}, yielding
\beq
\label{KretThinGen}
R_{\mu\nu\al\be}^2 \approx \frac{3 G^2 M^2 \left( S - S_2 \right)^2  \beta^4}{r^2}  \,   ,
\eeq
where $\beta^2 \equiv 1 - t^2 r^{-2}$ ranges between 0 and 1 outside the shell. As the collapse proceeds, the Kretschmann scalar diverges\footnote{Note, however, that the divergence is less strong than in GR, for which $R_{\mu\nu\al\be}^2 \sim r^{-6}$.} for $r\rightarrow 0$. This very same 
behavior occurs in the nonlocal ghost-free gravity~\cite{Frolov:Exp}, and it was previously verified to occur also in the particular case of polynomial models with simple poles~\cite{Frolov:Poly}. Actually, in view of these results on similar models it is natural to expect the nonregularity of the Kretschmann invariant, as in these cases the function $F(z)$ has the same linear dependence on $z$ for small arguments. As pointed out in~\cite{Frolov:Exp,Frolov:Poly} this singularity of  $R_{\mu\nu\al\be}^2$ is associated to the nonphysical assumption of an infinitesimally thin shell. The physical imploding shell must have finite thickness, which tends to regularize the curvature (see Sec.~\ref{Sec.5}).

It is also instructive to recall that for the fourth-order gravity, \textit{i.e.}, $\mathcal{N} = 1$, one gets
\beq
q(z) = - \frac{z}{4} (\ln z  + q_1 )m_1^2 + O(z^2)  \, ,
\eeq
around $z=0$, where $q_1 \equiv 2 \gamma - 2 \ln 2 - 1 + \ln m_1^2$. This gives~\cite{Frolov:Poly}
\beq
\label{KretThin4th}
R_{\mu\nu\al\be}^2  \approx \frac{3 G^2 M^2 m_1^4\beta^4}{r^2} \Big\lbrace q_1^2 + 2(1 +  q_1 )\left[ 1+\ln (\be^2 r^2)\right] 
 +  \left[ \ln (\be^2 r^2)\right] ^2 \Big\rbrace   \,  ,
\eeq
which diverges more rapidly (\textit{c.f.} Eq.~\eqref{KretThinGen}) as the collapse proceeds and $r\rightarrow 0$.

\section{Thick null shell collapse}
\label{Sec.5}

In the linear regime one can build the metric associated to thick null shell by superposing a set of $\delta$-shells collapsing to the same spatial point $O$, which we take as origin of the coordinate system. Of course, there are infinite possibilities of distributing the total energy of the shell throughout its thickness. Since our goal is to show that a nonsingular source regularizes the Kretschmann scalar in the polynomial gravity (and ameliorates the divergence for the fourth-order model), we choose the most simple profile by assuming that the density $\rho(t)$ at $r=0$ remains constant during the collapse, being null before/after it. Such a definition of the energy flux passing at $O$ suffices to determine the density profile of the shell, insomuch as each element of the fluid moves at the speed of the light and no self-interaction is considered inside the shell. Therefore, for a shell with total mass $M$ and thickness (or duration) $\tau$,
\beq
\rho(t) = \left\{ 
\begin{array}{l l}
0 \, ,  &  \text{if $\,|t|>\tau/2\,$,}\\
M/\tau \, , &  \text{if $\,-\tau/2<t<\tau/2\,$,}\\
\end{array} \right .
\eeq
where we set $t=0$ as the moment when half of the total mass crosses $O$. The corresponding metric perturbation can be obtained by averaging the metric~\eqref{dhr} of the thin null shells with respect to the density $\rho$~\cite{Frolov:Exp},
\beq
\n{llrr}
\langle \langle dh^2\rangle \rangle (t,r)
\,=\, 
\int dt' \rho(t') \langle dh^2 \rangle(t-t',r)
\,.
\eeq

The collapse of a thick null shell defines specific spacetime domains (see, \textit{e.g.},~\cite{Frolov:Exp}). In the present work we restrict considerations to the domain 
$I$ near $t = r = 0$, where (and when) the shell assumes its highest density---favoring the mini black hole formation and the emergence of singularities. This domain is characterized by the intersection of the in-coming and the out-coming fluxes of null fluid, and it is formally defined by the locus of the spacetime points for which $r + \vert t \vert < \tau/2$. Moreover, the metric is stationary inside $I$, for the energy density is constant.
Taking into account that only the $\delta$-layers which cross $O$ at times $t^\prime \in (t-r,t+r)$ contribute to the field inside this domain, it is not difficult to verify that Eq.~\eq{llrr} yields~\cite{Frolov:Exp}
\beq
\label{Met_Thick}
\langle \langle dh^2\rangle \rangle
= -\frac{2G M}{\tau r}\left[  J_0  dt^2 + J_2 \frac{dr^2}{r^2}
+  \frac{1}{2} \left( J_0 r^2  -J_2 \right) d\Omega^2
\right] ,
\quad r + \vert t \vert < \frac{\tau}{2} \, ,
\eeq
where we defined
\beq
\label{J-def}
J_n(r) \equiv \int_{-r}^{r} dx \ x^n\  F(r^2-x^2).
\eeq

Particularizing this solution for gravity models with six or more derivatives in the action, we substitute the expression~\eqref{F_expand2} for $F(z)$ around $z=0$. It follows
\beq
\label{J0J2}
J_0(r) = 2 c^\prime r - \frac{r^3}{3} (S - S_2) + O(r^5) \, , 
\qquad
J_2(r) = \frac{2 c^\prime r^3}{3}  - \frac{r^5}{15} (S - S_2) + O(r^7) \, .
\eeq
The Kretschmann scalar associated to this solution is
\beq
R_{\mu\nu\al\be}^2 \, = \, \frac{32 G^2 M^2 \left( S - S_2 \right)^2 }{3 \tau^2}  \, + \, O(r^2) \ ,
\eeq
which is regular at $r=0$, as anticipated. It is worthwhile to mention that the nonsingularity of the source is not enough, by itself, to guarantee the regularity of the curvature. In fact, $F(z) \sim \ln z$ in GR, which gives $R_{\mu\nu\al\be}^2 \sim r^{-4}$ for the collapsing thick null shell. Also, the presence of the term $z \ln z$ in the small-$z$ expansion of $F(z)$ could yield logarithmic divergences in the Kretschmann scalar. Such singularity was considered in~\cite{Frolov:Poly} as a possibility for general higher-derivative polynomial gravity (see Eq.~\eqref{F_expand1}). Nonetheless, it only occurs for the models with four derivatives in the spin-2 sector, since for nontrivial polynomial theories there is the relation $S_1 = S_2$ which regularizes the potential $\chi$.

Explicitly, the Kretschmann scalar for a collapsing thick null shell in the fourth-derivative gravity follows from~\eqref{F_4th} and reads~\cite{Frolov:Poly}
\beq
R_{\mu\nu\al\be}^2 = \frac{32 G^2 M^2 m_1^4 }{27 \tau^2} \left[ 5 + 9c^2 + 36 c \ln r + 36 (\ln r)^2\right] 
 +  O(r^2) \ ,
\eeq
with $c \equiv 2 \gamma - 2 + \ln m_1^2$.
The origin of this singularity can be traced back to the nonrelativistic limit. Indeed, in~\cite{Frolov:Poly} it was shown that, for polynomial theories with simple poles, the nonregularity of the  potential $\chi$ implied in a singular Kretschmann scalar for the collapsing thick null shell.

We have seen that the divergences are softened when a $\delta$-shell is substituted by a thick shell. It is therefore natural to expect the existence of a mass gap to the formation of mini black holes for a collapsing thick null shell too. For the sake of completeness, we calculate the invariant $g(r)$ on the the domain $I$ for the solution~\eqref{J0J2}, which reads
\beq
g(r) \, = \,  1 + \frac{2 G M (S - S_2) r^2}{3\tau} \, + \, O(r^4) \, .
\eeq
Since $r < \tau$ on $I$, it follows that
\beq
\frac{2 G M \vert S - S_2 \vert r^2}{3\tau}  <   \frac{2 G M \vert S - S_2 \vert \tau}{3} \, .
\eeq
Hence, for a given $\tau$ it is also possible to avoid the existence of an apparent horizon inside  $I$ provided that the mass $M$ is sufficiently small.

\section{Summary and discussion}
\label{Sec.6}

Let us summarize the results obtained. We derived the solutions for
the Newtonian potentials associated to a pointlike mass in a general
polynomial higher-derivative gravity, \textit{i.e.}, allowing the presence of complex and
degenerate poles (with arbitrary order) on the propagator. This includes the classes of
(super)renormalizable theories and Lee-Wick gravity models. 
It was verified, in agreement to~\cite{Newton-BLG}, that the metric potentials remains finite in $r=0$ provided that there is at least one massive mode in each spin-$2$ and spin-$0$ sectors. This is not a sufficient condition, however, 
to ensure the regularity of the solution, because there can be singularities in the curvatures.

Indeed, since the 1970s it is known that Stelle's fourth-order gravity possesses curvature singularities  in the linear regime~\cite{Stelle78}. On the other hand, there were evidences that such singularities would be regularized in the models which contain more than four derivatives in the action~\cite{Modesto-LWBH,Holdom}. 
Using the expressions~\eqref{Phi-Gen} and~\eqref{Psi-Gen} derived in Sec.~\ref{Sec.2} for the nonrelativistic potentials $\ph$ and $\psi$, we showed explicitly that in a generic polynomial gravity with more than four derivatives in both scalar and spin-2 sectors the curvatures remain finite at the origin. This result completely characterizes the class of local higher-derivative gravities which have a regular Newtonian limit.

In the ensuing part of the paper we considered the dynamical process of the spherically symmetric collapse of null shells in linearized  higher-derivative polynomial gravities. Here we generalized the discussion carried out in~\cite{Frolov:Poly} to include the possibility of degenerate poles.
If one allows the shell to have a certain thickness, then the Kretschmann invariant becomes finite during the collapse provided that the model has at least six derivatives in its spin-2 sector.
This observation on the regularity of the metric of the thick shell is a refinement of the result derived in~\cite{Frolov:Poly}. Indeed, the logarithmic divergences of the Kretschmann scalar which in principle could occur in polynomial theories are actually ruled out in most of the cases, due to a specific algebraic relation between the poles of the propagator. Only in the fourth-order gravity these logarithmic divergences are possible.
Finally, we have shown that, like in the case of polynomial gravities with simple poles in the propagator~\cite{Frolov:Poly}, there exists a mass gap for the mini black hole formation also in the models with higher-order poles.

 With the results obtained in the present work it is possible to observe some similarities between the nonlocal (ghost-free) higher-derivative gravity and the local (polynomial) models with more than four derivatives. First, in both theories there is the cancellation of the Newtonian singularity of the metric potentials associated to a $\delta$-source \cite{Tseytlin95,Modesto12,Maz12,Newton-MNS,Newton-BLG,EKM16}. 
Second, it is known that in the nonrelativistic limit there is a class of nonlocal gravities that have a regular solution for the field generated by a pointlike source \cite{Head-On,Buoninfante:2018b,Buoninfante:2018a}. Our results show that in a generic polynomial higher-derivative gravity with more than four derivatives in each sector the Newtonian limit is regular too. (Actually, using the description of effective sources presented in~\cite{BreTib2} it is possible to deduce the regularity of some nonlocal theories from the comparison with a sequence of sources associated to the local models.)
A third similarity is the regularity of the metric of the collapsing shell. In fact, if a thin shell is considered, nonpolynomial and nontrivial polynomial theories have a Kretschmann scalar which diverges quadratically for small $r$~\cite{Frolov:Exp,Frolov:Poly}. This is, however, the consequence of the nonphysical assumption of an infinitesimally thin shell. If the shell has some thickness, then in both theories the Kretschmann invariant becomes finite during the collapse. This happens, again, because the leading term in the expansion of Eq.~\eqref{F_DR} around $z=0$ is the linear one, just like what occurs in the nonlocal ghost-free gravity (see~\cite{Frolov:Exp,Frolov:Poly}). Solely in the fourth-order gravity the divergences in the Kretschmann invariant are possible, a situation analogous to what happens in the Newtonian limit.
Moreover, in both theories there is a mass gap for the mini black hole formation. Indeed, this feature is present in any higher-derivative model with an arbitrary number of derivatives in the spin-2 sector~\cite{Frolov:Exp,Frolov:Poly,Frolov:Weyl}, since in these theories there is a new mass scale.
These four connections between polynomial and ghost-free gravity theories can be supportive of the view that the nonlocal models may be considered as the limit of a theory with an infinite amount of complex poles hidden at the infinity~\cite{CountGhost}. In this sense, it is useful to notice that many good regularity properties of the nonlocal gravity~\cite{Modesto12,Buoninfante:2018b,Buoninfante:2018a} can be achieved without the need of losing locality at the classical level, and may be common to models with at least six derivatives. Further discussion on the similarities between local and nonlocal models are carried out in the parallel work~\cite{BreTib2}.

All the results which were mentioned above have been obtained in the linear approximation. The most interesting question is whether there can be nonsingular solutions in the full nonlinear regime of polynomial gravity theories. The first step in this direction was done within the fourth-order gravity in Ref.~\cite{Stelle78}, where the asymptotic analysis of the static field equations near the origin was carried out via the Frobenius technique. It was shown the existence of three families of solutions: a set of nonsingular solutions, and two sets of singular ones---one of them containing the Schwarzschild solution. The presence of the Schwarzschild solution is expected, because by means of the Gauss-Bonnet relation
\beq
\n{2gb2}
\int d^4 x \sqrt{-g}\, E \,=\, \mbox{total derivative},
\eeq
where $E = R_{\mu\nu\al\be}^2 - 4 R_{\mu\nu}^2 + R^2$,
it is possible to completely remove the Riemann-squared term of the action, and it is clear that any vacuum solution of the Einstein
equations ($R_{\mu\nu} = 0$) is also a solution of the fourth-order gravity~\cite{Stelle78,Frolov:2009qu}. Nonetheless, in this model the Schwarzschild solution is not coupled to a positive-definite matter source~\cite{Stelle78}.
 More recently, some new aspects of the nonlinear static spherically solutions in fourth-order gravity were considered in~\cite{Holdom,Stelle15PRL,Stelle15PRD} by means of numerical methods.
In particular, it was studied what happens when the asymptotic solutions in strong-field regime near $r=0$ are linked with the weak-field 
solution at large $r$ in the form of a combination of Newton and Yukawa potentials---such a potential is the particular case of our general result, Eqs.~\eqref{Phi-Gen} and \eqref{Psi-Gen}. In summary, the result is that for a $\de$-like source the solution has no horizon and falls to a timelike singularity at $r=0$. Actually, the presence of the singularity in this solution is expected in view of the fact that $R_{\mu\nu\al\be}^2$ diverges yet at the linear regime. Moreover, the absence of horizon in the full fourth-order model is guaranteed by a general theorem~\cite{Stelle15PRL,Stelle15PRD,Nelson:2010ig}; and only the particular theories where the $R^2$-term is excluded from the action could have horizons. 

In what concerns the theories with derivatives higher than fourth, in Ref.~\cite{Holdom} the asymptotic solutions near $r=0$ were studied by the Frobenius series expansion method in models with up to $10$ derivatives in the action. It was shown that there is no Schwarzschild-like solutions, or other ones with singularity. Only the nonsingular solutions remain in the static spherically symmetric case for sixth- and higher-order theories\footnote{We point out, however, that the method based on the expansion in Frobenius series around $r=0$ is not sufficient to rule out the existence of singularities, as there may be solutions with a violent singularity which does not admit such representation at the origin. Also, it is possible to have solutions with singularities at a finite radius.}. The nonexistence of the exact Schwarzschild solution is due to the absence of the Gauss-Bonnet relation for the higher-order terms. The analogue relation~\eqref{gb} is insufficient to eliminate the effect of the Riemann-squared terms in the nonlinear regime, since
$O(R^3)$ structures still remain. Also, the nature of these nonsingular solutions implies that the complete solutions with large $r$ behavior given by Eqs.~\eqref{Phi-Gen} and \eqref{Psi-Gen} must have no horizon or an even number of horizons. Another interesting result of Ref.~\cite{Holdom} is the necessity of theories with six or more derivatives to the possible elimination of the de~Sitter-like horizons.

The results of the present work, in light of~\cite{Holdom}, bring more motivations for further investigation of the spherically symmetric static solutions in the full nonlinear regime for the polynomial theories with more than four derivatives. It would also be interesting to know whether in these theories there is some kind of no-horizon theorem, and we expect to revisit this issue in the future. In case of a positive answer, the complicated numerical search of solutions might be simplified.

\appendix
\section*{Appendix A: Useful identities with the coefficients $A_{i,j}$}
\label{Appendix}

Let $a(z)$ be a polynomial function of degree $\mathcal{N} \geq 1$ which satisfies $a(0)=1$.
The quantities $A_{i,j}$ defined by~\eqref{Aij} are related to the coefficients $a_{i,j}$ of the partial fraction expansion of
\beq
\label{PartialFrac}
-\frac{1}{\xi a(-\xi)} \, = -\frac{1}{\xi} + \sum_{i=1}^N \sum_{j=1}^{\alpha_i} a_{i,j}  \frac{1}{(\xi + m_i^2)^{j}} \, .
\eeq
In fact, $\, a_{i,j} =  A_{i,j}(j-1)! \,$ and, in particular, $\, A_{i,1} = a_{i,1}\,$ and  $A_{i,2} = a_{i,2}$. Proceeding the regrouping of the \textit{r.h.s.} into a single fraction one obtains
%
\beq
\label{PartialFrac2}
-\frac{1}{\xi a(-\xi)} = \frac{-\prod_{i=1}^{N}(\xi + m_i^2)^{\alpha_i}
 + \xi \sum_{i=1}^N \sum_{j=1}^{\alpha_i} a_{i,j}(\xi + m_i^2)^{\alpha_i - j} \prod_{k \neq i}(\xi + m_k^2)^{\alpha_k}}{\xi \prod_{i=1}^{N}(\xi + m_i^2)^{\alpha_i}}  \, .
\eeq

Comparing the numerators of the fractions above order by order in $\xi$, one obtains for the highest order term ($\mathcal{N} = \sum_i \alpha_i$)
\beq
0 = \left( - 1 + \sum_{i=1}^N a_{i,1} \right) \xi^{\mathcal{N}} \, ,
\eeq
whence
\beq
\label{SumAi1}
\sum_{i=1}^N A_{i,1} = 1 \, .
\eeq
The substitution of this result into~\eqref{Pot.r=0} shows that the Newtonian singularity is canceled
in general higher-derivative models.

Now, let us assume that $\mathcal{N} \geq 2$. Comparing both sides of~\eqref{PartialFrac2} for the term proportional to $\xi^{\mathcal{N}-1}$ one obtains
\beq
\sum_{i=1}^N \Big[ - m_i^2 \alpha_i + A_{i,1} \Big(  m_i^2 (\alpha_i - 1) +  \sum_{j \neq i}  m_j^2 \alpha_j \Big) + A_{i,2} \Big] 
  = 0 \, .
\eeq
Since, for a given $i$,
\beq
\sum_{j \neq i}  m_j^2 \alpha_j = - m_i^2 \alpha_i + \sum_{j}  m_j^2 \alpha_j \, ,
\eeq
and using~\eqref{SumAi1}, it follows that
\beq
\label{AppS1S2}
\sum_{i=1}^N  A_{i,2} = \sum_{i=1}^N A_{i,1} m_i^2 \, .
\eeq
In terms of the definitions in the Eq.~\eqref{S_def}, the identity above reads $S_2 = S_1$. We recall that this relation is valid only if $\mathcal{N} \geq 2$. The case $\mathcal{N} = 1$ implies in $S_1 = m_1^2$ and $S_2=0$.

\section*{Acknowledgements}

The authors are thankful to I.L. Shapiro for the useful discussions.
B.L.G is grateful to CNPq--Brazil for supporting his Ph.D. project.
T.P.N. wishes to acknowledge 
CAPES for the support through the PNPD program.
B.L.G. is grateful to the Department of Physics of the 
Universidade Federal de Juiz de Fora, where this work was 
carried out, for the warm hospitality during his visit.


\end{document}